\documentclass[sigconf,nonacm]{acmart}

\renewcommand\footnotetextcopyrightpermission[1]{}

\usepackage{booktabs}
\usepackage{tabularx}
\usepackage{listings}
\usepackage{xcolor}
\usepackage{amsmath}

\setcopyright{none}

\begin{document}

\title{Type-Checked Compliance: Deterministic Guardrails for Agentic Financial Systems Using Lean 4 Theorem Proving}

\author{Devakh Rashie}
\thanks{Code, implementation, and live demo available at \url{https://github.com/arkanemystic/lean-agent-protocol} and \url{https://axiom.devrashie.space}.}
\thanks{Suggested Citation: Devakh Rashie, Veda Rashi. 2026. Type-Checked Compliance: Deterministic Guardrails for Agentic Financial Systems Using Lean 4 Theorem Proving. Preprint available at \url{https://devrashie.space}.}
\affiliation{%
  \institution{Independent Researcher}
  \city{Centreville}
  \state{Virginia}
  \country{USA}
}
\email{devrashie@gmail.com}

\author{Veda Rashi}
\affiliation{%
  \institution{Thomas Jefferson High School for Science and Technology}
  \city{Alexandria}
  \state{VA}
  \country{USA}
}
\email{vedarashi21@gmail.com}

\begin{abstract}
The rapid evolution of autonomous, agentic artificial intelligence within financial services has introduced
an existential architectural crisis: large language models (LLMs) are probabilistic, non-deterministic
systems operating in domains that demand absolute, mathematically verifiable compliance guarantees.
Existing guardrail solutions—including NVIDIA NeMo Guardrails and Guardrails AI—rely on
probabilistic classifiers and syntactic validators that are fundamentally inadequate for enforcing complex
multi-variable regulatory constraints mandated by the SEC, FINRA, and OCC. This paper presents the
\emph{Lean-Agent Protocol}, a formal-verification-based AI guardrail platform that leverages the
Aristotle neural-symbolic model developed by Harmonic AI to auto-formalize institutional policies into
Lean~4 code. Every proposed agentic action is treated as a mathematical conjecture: execution is
permitted if and only if the Lean~4 kernel proves that the action satisfies pre-compiled regulatory axioms.
This architecture provides cryptographic-level compliance certainty at microsecond latency, directly
satisfying SEC Rule 15c3-5, OCC Bulletin 2011-12, FINRA Rule 3110, and CFPB explainability
mandates. A three-phase implementation roadmap from shadow verification through enterprise-scale
deployment is provided.
\end{abstract}

\keywords{formal verification, AI guardrails, Lean 4, agentic AI, financial regulation, theorem proving,
neural-symbolic AI, SEC compliance, formal methods}

\maketitle

\section{Introduction: Deterministic Governance in Agentic AI}

The rapid evolution of artificial intelligence within the financial services sector has catalyzed a
transition from isolated, predictive analytics and generative text models to fully autonomous, agentic
systems. These frameworks possess the capability to interpret complex user intent, dynamically
synthesize contextual data, formulate multi-step execution plans, and autonomously interact with
external environments through application programming interfaces (APIs) and tool-calling
mechanisms~\cite{ibm2026agents}. While this paradigm shift offers massive operational efficiencies in
algorithmic trading, dynamic risk assessment, and autonomous portfolio management, it simultaneously
introduces an existential architectural crisis regarding systemic reliability and regulatory
compliance~\cite{finra2026aiapps}.

Traditional enterprise automation relies on deterministic systems—such as robotic process automation
(RPA) bots and hard-coded workflow engines—that execute predefined logical paths with mathematical
precision~\cite{ibm2026agents}. If a condition is not explicitly defined, execution halts, bounding
operational risk. Conversely, LLMs and the agentic architectures built upon them operate
probabilistically~\cite{purplesec2026}. They navigate continuous, high-dimensional vector spaces to
generate responses that are \emph{statistically likely} rather than mathematically certain~\cite{rainbird2025}.
This stochastic nature renders them highly adaptable to ambiguous inputs but inherently susceptible to
hallucinations, context drift, and adversarial prompt injections~\cite{purplesec2026}. In highly regulated
domains such as global financial markets—where a single algorithmic deviation can precipitate
catastrophic capital depletion or systemic market instability—probabilistic execution without rigid
constraints is architecturally and legally untenable~\cite{finra2021marketaccess}.

The contemporary industry response has been the deployment of \emph{AI Guardrails}—secondary
software layers designed to filter, monitor, and control LLM inputs and outputs~\cite{ajayverma2026guardrails}.
However, the vast majority of existing guardrail solutions rely heavily on probabilistic classifiers, vector
similarity searches, or syntactic schema validation~\cite{nvidiacolang2026}. While these tools excel at
content moderation, they are fundamentally inadequate for enforcing the complex, multi-variable
regulatory constraints demanded by the SEC, FINRA, and OCC~\cite{finra2026aiapps}.

To reconcile the agility of agentic AI with the uncompromising mandates of financial regulation, a
paradigm shift toward formal verification is required. This paper delineates the technical blueprint,
regulatory feasibility, and implementation roadmap for the \textbf{Lean-Agent Protocol}. By leveraging
the Aristotle model~\cite{aristotle2025arxiv} developed by Harmonic AI, this platform translates natural
language institutional policies into the dependently typed formal programming language Lean~4~\cite{lean4lang}.
Acting as a deterministic gateway, the Lean-Agent Protocol treats every proposed agentic action as a
mathematical conjecture: execution is permitted if and only if the Lean~4 kernel can definitively prove
that the action satisfies the pre-compiled regulatory axioms. V.~Rashi led the formal mapping of
SEC Rule 15c3-5 and FINRA Rule 3110 regulatory mandates into the Lean-Agent logical framework.

\section{Technical Integration and Workflow Dynamics}

\subsection{The Aristotle Model: Neural-Symbolic Architecture}

At the foundational core of the Lean-Agent Protocol is the Aristotle model, engineered by the Silicon
Valley-based startup Harmonic AI~\cite{aristotleapi2026}. Aristotle represents a structural departure from
purely autoregressive LLMs by pioneering a neural-symbolic methodology designed specifically for
automated theorem proving and the generation of formally verified, hallucination-free
reasoning~\cite{aristotleapi2026}. The system gained international prominence by achieving
gold-medal-equivalent performance on the 2025 International Mathematical Olympiad (IMO),
successfully generating machine-verifiable solutions in Lean~4 for five out of the six competition
problems~\cite{aristotle2025arxiv}.

The architecture of Aristotle integrates three distinct but synergistic components: (1) a continuous Lean
proof search system, (2) an informal reasoning engine dedicated to generating and formalizing
intermediary lemmas, and (3) a specialized geometry solver~\cite{aristotle2025arxiv}. Unlike traditional
models that predict the next token in isolation, Aristotle utilizes the Lean~4 compiler as an infallible,
high-speed reward signal within a reinforcement learning loop~\cite{aristotle2025arxiv}. This mathematical
grounding forces the model to continuously validate its internal reasoning against the strict rules of the
Calculus of Inductive Constructions (CIC)~\cite{lean4lean2024}.

The primary capability that enables the Lean-Agent Protocol is Aristotle's proficiency in
\emph{auto-formalization}—the automated translation of unstructured, natural language concepts into
rigorous formal code~\cite{autoformalization2024}. During its training and deployment, Aristotle has made
novel contributions to Mathlib (Lean's comprehensive mathematical library), formalized proofs for
long-standing open conjectures including Erd\H{o}s Problems 1026 and 728, and identified subtle logical
errors within established academic textbooks~\cite{aristotleapi2026}.

Crucially, Aristotle is engineered to handle ``epistemic gaps'' and incomplete proof skeletons.
When translating a complex financial policy, the initial mapping may leave certain logical steps
unproven—denoted in Lean by the \texttt{sorry} tactic~\cite{aristotleapi2026}. Aristotle possesses the
capability to autonomously deduce and fill multiple \texttt{sorry} markers within a single theorem,
effectively bridging the gap between high-level regulatory directives and the granular logic gates
required for machine execution~\cite{aristotleapi2026}.

\subsection{Translation Error Rates and Handling of Complex Logical Constraints}

A critical concern when translating natural language policies into executable code is the fidelity of the
translation. The Aristotle model mitigates this risk not by achieving perfect initial generation, but by
leveraging the deterministic constraints of the Lean environment to execute autonomous
self-repair~\cite{aristotle2025arxiv}.

In the formalization of complex theorems, Aristotle's informal reasoning engine occasionally generates
text containing subtle semantic confusions—for example, conflating the mathematical definitions of
``strictly decreasing sequences,'' ``weakly decreasing sequences,'' and sequences that merely decrease at
some point~\cite{aristotle2025arxiv}. In a standard generative AI pipeline, this semantic error would pass
through unnoticed. However, within the Aristotle architecture, the informal proof is immediately
submitted to the Lean~4 kernel for formalization~\cite{aristotle2025arxiv}. The Lean compiler
deterministically rejects the flawed logic and generates a highly specific compiler error, often including
explicit counterexamples~\cite{aristotleapi2026}. Aristotle ingests this error state and autonomously
repairs the logical gaps, adjusting its approach until the Lean compiler returns a successful
verification~\cite{aristotle2025arxiv}. The final output is therefore guaranteed to be correct relative to the
formalized problem statement, effectively neutralizing the risk of ``hallucinated compliance''~\cite{harmonic2026hacker}.

\subsection{Architectural Design: The Deterministic Request-Response Loop}

To operationalize the Aristotle engine as a secure guardrail, the Lean-Agent Protocol employs a
state-machine-driven Request-Response loop. This architecture intercepts the probabilistic output of a
primary trading agent and forces it through a deterministic verification bottleneck prior to API execution.
The architectural flow is divided into two distinct temporal phases:

\subsubsection{Phase 1: Asynchronous Policy Configuration}
\begin{enumerate}
  \item \textbf{Natural Language Ingestion.} Compliance officers author institutional trading limits,
  risk parameters, and regulatory mandates in standard English (e.g., \emph{``Do not execute trades
  exceeding 10\% of the firm's available daily capital.''}).
  \item \textbf{Aristotle Translation.} The Aristotle API processes these documents, utilizing its
  informal reasoning engine to parse the text and its proof search system to auto-formalize the
  constraints into Lean~4 code.
  \item \textbf{Compilation and Storage.} The generated code is compiled by the Lean kernel. Any
  logical inconsistencies are iteratively repaired by Aristotle. Once successfully compiled, the
  resulting axioms, theorems, and definitions are stored in an immutable repository known as the
  \emph{Policy Environment}.
\end{enumerate}

\subsubsection{Phase 2: Synchronous Runtime Authorization (The Agentic Loop)}
\begin{enumerate}
  \item \textbf{Intent Generation.} The primary Agentic AI (e.g., a portfolio rebalancing agent)
  analyzes real-time market data and generates an intent to execute a specific API tool (e.g.,
  \texttt{execute\_trade(symbol="AAPL", volume=50000, type="market")}).
  \item \textbf{Orchestrator Interception.} The Orchestrator node intercepts this API call before
  it reaches the execution environment. It tokenizes the request, extracting parameters and
  current systemic state variables (available capital, recent volatility indices)~\cite{nvidiacolang2026}.
  \item \textbf{Conjecture Formulation.} The Orchestrator maps these parameters against the
  Lean-verified theorems in the Policy Environment, formatting the agent's proposed action as a
  formal mathematical conjecture~\cite{axprover2025}.
  \item \textbf{Lean Kernel Verification.} The conjecture is submitted to the Lean~4 type-checker,
  which attempts to prove that the proposed action satisfies all constraints within the
  Policy Environment.
  \item \textbf{Deterministic Actuation.}
  \begin{itemize}
    \item \textbf{State A (``Proven'' / Allowed):} If the kernel successfully verifies the proof,
    it returns a binary \texttt{True}. The Gateway API unlocks and the original API call is
    routed to the execution layer.
    \item \textbf{State B (``Refuted'' / Blocked):} If the kernel cannot verify the proof (e.g.,
    the trade volume exceeds the capital theorem), it returns a binary \texttt{False}.
    The action is definitively blocked, and the system generates a formal error trace
    for the audit log~\cite{deterministic2026rulebricks}.
  \end{itemize}
\end{enumerate}

\subsection{Real-Time Computational Latency in the Agentic Loop}

A persistent historical critique of formal verification methods is their high computational overhead,
which traditionally rendered them unsuitable for low-latency, real-time applications such as
high-frequency trading (HFT)~\cite{democratizing2025}. However, architectural advancements within
Lean~4, specifically the decoupling of proof generation from proof checking, fundamentally resolve
this latency bottleneck.

In the Lean-Agent Protocol, the computationally intensive process of generating proofs and formalizing
policies occurs entirely asynchronously during the configuration phase~\cite{aristotleapi2026}. During
synchronous runtime execution, the system only requires the Lean~4 kernel to perform type-checking and
proof verification against pre-compiled binaries~\cite{lean4lean2024}. The Lean trusted kernel is heavily
optimized, relying on minimal primitive operations written in highly performant C++~\cite{lean4lean2024}.

Empirical benchmarking derived from Amazon Web Services' (AWS) deployment of the Cedar
authorization policy language—which is formally verified using Lean~4—demonstrates the extreme
efficiency of this approach~\cite{awslean2026}. In differential testing environments, evaluating a
formalized access control input against the Lean model requires an average of only \textbf{5
microseconds}~\cite{awslean2026}. This execution speed is notably faster than the equivalent evaluation
in highly optimized Rust code, which averages 7 microseconds per test case~\cite{awslean2026}.

Furthermore, recent updates to the Lean~4 ecosystem have introduced advanced automated reasoning
tactics, such as \texttt{grind}, which integrate SMT (Satisfiability Modulo Theories) solving
capabilities directly into the prover~\cite{lean4230}. Consequently, when the Orchestrator submits a
simple arithmetic constraint (e.g., checking if $\mathit{Trade\_Value} \leq \mathit{Capital\_Limit}$),
the Lean kernel resolves the statement with sub-millisecond latency~\cite{apollo2025}. This completely
removes formal verification as a bottleneck in the execution critical path~\cite{apollo2025}.

\section{Financial Regulatory and Compliance Mapping}

\subsection{SEC Rule 15c3-5: The Market Access Rule and Hard Guardrails}
\label{sec:sec15c3}

The deployment of autonomous trading agents introduces severe risks to market integrity, prompting
intense regulatory scrutiny. Following instances of devastating algorithmic failures—such as the 1987
market crash and the Knight Capital incident in 2012 (where an unchecked algorithm lost \$440 million
in 45 minutes)—the Securities and Exchange Commission (SEC) implemented Rule 15c3-5, known as
the Market Access Rule~\cite{nasdaq2025algofailures}.

SEC Rule 15c3-5 mandates that broker-dealers providing market access must ``establish, document, and
maintain a system of risk management controls and supervisory procedures reasonably designed to
manage the financial, regulatory, and other risks''~\cite{finra2025marketaccess}. Specifically, the rule
requires the implementation of pre-trade controls that: (1)~prevent the entry of orders that exceed
appropriate pre-set credit or capital thresholds; and (2)~prevent the entry of erroneous orders by
rejecting requests that exceed appropriate price or size parameters~\cite{sec15c3faq}.

Crucially, the SEC dictates that these controls must be under the ``direct and exclusive control'' of the
broker-dealer~\cite{wilmerhale15c3}. Deploying a secondary, probabilistic LLM as a safety filter
constitutes a severe regulatory vulnerability~\cite{purplesec2026}. LLMs are fundamentally non-deterministic;
their behavior is governed by stochastic sampling across continuous vector representations~\cite{arxivagentssecurity2025}.
A probabilistic filter might successfully block 99.9\% of non-compliant trades, but the inherent
mathematical possibility of an evasion means the firm cannot guarantee absolute compliance, thereby
failing the standard of ``exclusive control''~\cite{purplesec2026}.

The Lean-Agent Protocol directly satisfies the rigid mandates of SEC Rule 15c3-5 by enforcing
\emph{hard} deterministic guardrails~\cite{moveo2026deterministic}. Because the capital thresholds and
price parameters are formalized as immutable Lean~4 axioms, the AI agent's execution layer is
cryptographically bound by mathematical logic~\cite{probabilisticlogic2026}. The verification kernel
does not interpret intent or estimate probability; it computes binary truth. If an agent proposes a trade
where the size parameter mathematically exceeds the stated limit, the proof fails compilation and the
order is deterministically rejected at the gateway level~\cite{deterministic2026rulebricks}. This framework
provides regulators with verifiable, mathematical proof that erroneous orders cannot physically reach
the market, fully aligning agentic automation with the strictest interpretations of the Market Access Rule.

\subsection{OCC Bulletin 2011-12 and FINRA Rule 3110: Supervisory Obligations}
\label{sec:finra3110}

Beyond direct market access, the Office of the Comptroller of the Currency (OCC) and the Financial
Industry Regulatory Authority (FINRA) impose broad requirements for organizational risk management
and the supervision of automated systems. OCC Bulletin 2011-12 (SR 11-7) established the foundational
framework for Model Risk Management (MRM), requiring robust model development, implementation,
continuous validation, and sound governance~\cite{occ201112}.

Traditional MRM frameworks were designed under the assumption that computational models were
deterministic entities with easily interpretable inputs and predictable outputs~\cite{moodys2026mrm}.
The introduction of generative AI and autonomous agents completely fractures these assumptions,
introducing high opacity, stochastic reasoning, and autonomous decision-making loops that bypass static
validation cycles~\cite{moodys2026mrm}. Concurrently, FINRA Rule 3110 requires member firms to
establish and maintain a system to supervise the activities of all associated persons and systems to
achieve compliance with securities laws~\cite{finra3110}.

In Regulatory Notice 24-09 and the 2026 Annual Regulatory Oversight Report, FINRA explicitly
clarified that its rules are technology-neutral; if a firm deploys Generative AI tools, those tools must be
supervised with the same rigor as human communications and traditional decision-making
systems~\cite{finra2026oversight}.

The Lean-Agent Protocol serves as the vital structural bridge between the opacity of LLMs and the
transparency required by FINRA and the OCC. By isolating the probabilistic ``thinking'' of the AI
agent from the deterministic ``doing'' of the execution API, institutions can foster innovation without
compromising governance. Because the execution of any agentic task is entirely dependent on clearing
the Lean kernel, the firm can demonstrate to OCC and FINRA examiners that the overarching
supervisory system remains firmly in human control, operating with 100\% predictability regardless
of the underlying LLM's stochastic output~\cite{moveo2026deterministic}.

\subsection{The Right to Explanation and Reverse Auto-Formalization for Audit Trails}

One of the most complex regulatory challenges facing the deployment of AI in financial services is the
legal mandate for algorithmic transparency, commonly referred to as the ``Right to
Explanation''~\cite{verityai2026audittrails}. Under the Equal Credit Opportunity Act (ECOA) and the
Fair Credit Reporting Act (FCRA), creditors are legally obligated to provide consumers with specific,
accurate, and easily understandable reasons when taking an adverse action~\cite{cfpb2023aidenials}.
The Consumer Financial Protection Bureau (CFPB) has issued explicit guidance stating that the use of
complex, ``black-box'' algorithms or AI models does not exempt institutions from these transparency
requirements~\cite{cfpb2023aidenials}. Similarly, in Europe, the GDPR and the newly enacted EU AI
Act mandate that individuals subject to automated decision-making have the right to obtain meaningful
explanations of the logic involved~\cite{euaiact2025}.

When the Lean-Agent Protocol blocks an AI agent's proposed action, the output of a failed Lean
compilation is a highly technical, dependently typed mathematical error trace. While this perfectly
enforces a deterministic block, presenting a formal logic compiler error to a consumer or non-technical
auditor is legally insufficient and violates plain-language disclosure
requirements~\cite{verityai2026audittrails}. To solve this, the Lean-Agent Protocol executes a process of
\emph{reverse auto-formalization} or ``back-translation,'' converting rigid mathematical failure states
back into compliant, natural language adverse action notices~\cite{verityai2026audittrails}.

This capability is modeled on advancements embodied in the Herald dataset and the NL2Lean
reinforcement learning framework~\cite{herald2024arxiv,nl2lean2025}. The Herald framework demonstrates
how static analysis tools can be used to extract structured metadata—including theorem declarations,
dependency relationships, and precise proof states—from Lean~4
environments~\cite{herald2024arxiv}. When a proof fails in the Lean-Agent gateway, the system isolates
the exact line of code and the specific logical tactic that triggered the rejection (e.g., identifying that
the agent's proposed action failed the $\mathit{Debt\_to\_Income} < 0.43$ axiom)~\cite{herald2024arxiv}.
Using an approach similar to NL2Lean, a specialized, fine-tuned translation model ingests this isolated
proof state and, utilizing a Retrieval-Augmented Generation (RAG) pipeline, generates a precise,
step-wise informal explanation of the failure~\cite{herald2024arxiv}. This ensures that the explanation
provided to the auditor or consumer is not a hallucinated rationale, but a direct, one-to-one translation
of the mathematical constraint that forced the denial~\cite{herald2024arxiv}.

\section{Security, Vulnerability Analysis, and Formal Sandboxing}

\subsection{Formal Methods for LLM Safety}

The prevailing methodology for securing Large Language Models relies on empirical techniques such as
Reinforcement Learning from Human Feedback (RLHF), constitutional AI alignment, and automated
red-teaming~\cite{verifytool2026}. While these techniques shape core behavioral tendencies, they are
fundamentally reactive and probabilistic. As LLMs are integrated into multi-agent systems with access
to APIs, databases, and execution environments, the attack surface expands exponentially—often
referred to as the ``lethal trifecta'' of tool access, natural language interfaces, and untrusted external
data~\cite{mozillaguardrails2026}.

The concept of Formal Methods for LLM Safety seeks to elevate AI security from empirical testing to
mathematical certainty~\cite{fusionllmformal2024}. By applying techniques traditionally reserved for
safety-critical hardware and aerospace engineering, formal methods guarantee that a system satisfies
specific functional and security properties across all possible state spaces~\cite{fusionllmformal2024}.
The Lean-Agent Protocol embodies this paradigm: rather than endlessly patching an LLM against every
conceivable adversarial input, the protocol assumes the LLM is compromised and instead secures the
execution perimeter using mathematically verified code generation and runtime evaluation~\cite{veriguard2025}.

\subsection{Vulnerability Analysis: Jailbreaking the Aristotle Translation Layer}

While the Lean~4 kernel provides absolute security at the execution level, the system's primary
vulnerability resides in the translation layer. If an adversary can manipulate the Aristotle model into
generating structurally valid but semantically malicious Lean code, they can effectively bypass the
policy constraints. This threat vector is explored in recent security literature under the terms
``logical jailbreaks'' and ``formalization drift.''

\subsubsection{Logical Jailbreaks}
Traditional alignment training conditions models to refuse clearly harmful natural language prompts.
However, adversaries utilize techniques like LogiBreak and QueryAttack to systematically translate
malicious intents into complex logical expressions, structured non-natural query languages, or
ciphers~\cite{logibreak2025}. By shifting the prompt's distribution away from the standard safety-alignment
training data while preserving the underlying semantic intent, these attacks successfully trick the model
into processing the request~\cite{queryattack2025acl}.

\subsubsection{Formalization Drift and Perturbation}
Attackers employ Sentence-Level Perturbations (SLPs)—such as rephrasing, adding irrelevant context,
or using synonym substitutions—to confuse the model's semantic mapping~\cite{llmlogictranslators2026}.
In LLM-based logical translators, this linguistic variation often leads to ``symbol drift'' or
``formalization drift''~\cite{llmlogictranslators2026}. If an attacker utilizes an Indirect Prompt Injection
that causes Aristotle to map a restricted action variable to a permitted Lean symbol, the Lean kernel will
successfully verify the proof, unwittingly authorizing a policy violation~\cite{llmlogictranslators2025}.

\subsubsection{Mitigation via Concept-Symbol Constraints}
To secure the translation layer, the Lean-Agent Protocol must enforce rigid structure on the Aristotle
formalization process. Drawing on frameworks like MenTaL (Mapping Equivalent Natural language To
Assigned Logic), the architecture must compel the LLM to explicitly build a concept-symbol mapping
table before initiating the translation into Lean~\cite{llmlogictranslators2025}. By algorithmically
linking varied linguistic expressions to a hard-coded registry of shared Lean~4 symbols, the system
mitigates symbol drift and ensures that adversarial synonyms cannot hijack the logic solver~\cite{llmlogictranslators2025}.
Furthermore, the Orchestrator node must aggressively sanitize context hydration, utilizing static parsing
to extract only necessary variables and completely stripping untrusted external text before the prompt
reaches the formalization engine~\cite{finosaigovernance}.

\subsection{Execution Sandboxing: The Strategic Imperative of WebAssembly (WASM)}

Because Agentic AI systems are designed to autonomously plan, write, and execute code, any code
generated by the agent must be inherently treated as a hostile payload~\cite{sandboxai2026}. Executing
untrusted code introduces severe security risks, including potential Remote Code Execution (RCE),
arbitrary file system manipulation, and sensitive data exfiltration~\cite{nvidiaagentic2026}.

Traditional containerization platforms such as Docker possess critical architectural flaws that render
them inadequate for securing highly untrusted AI execution~\cite{sandboxways2026}. Docker containers
share the host operating system's kernel. If an AI agent generates a payload containing a kernel exploit,
it can escape container boundaries, escalate privileges, and compromise the host machine and internal
networks~\cite{sandboxai2026}.

To achieve bulletproof isolation without sacrificing performance, the Lean-Agent Protocol mandates the
use of \textbf{WebAssembly (WASM)} as its execution substrate~\cite{sandboxways2026}. WASM is a
highly compact, portable binary instruction format designed to execute code at near-native speeds within
a mathematically secure sandbox~\cite{sandboxways2026}. The security guarantees of WebAssembly are
rooted in its architecture:

\begin{itemize}
  \item \textbf{Linear Memory Model.} WASM modules operate within a single, contiguous block of
  linear memory. Executing code has zero visibility into addresses outside this designated block,
  providing strict fault isolation from the host and other modules~\cite{wasmwrong2026}.
  \item \textbf{Protected Call Stack.} WASM utilizes a protected call stack logically separated from
  the module's heap memory. This renders control-flow hijacking mathematically
  impossible~\cite{wasmsecurity2024}.
  \item \textbf{Capability-Based Security (WASI).} By integrating the WebAssembly System Interface
  (WASI), the sandbox operates on a strict principle of least privilege. The WASM instance has
  zero default access to the host's filesystem, network sockets, or environment
  variables~\cite{sandboxways2026}.
\end{itemize}

By compiling both the Lean~4 verification environment and the agent's executable tools into WASM
binaries, the Lean-Agent Protocol establishes a zero-trust execution perimeter. This guarantees that even
if a sophisticated adversary successfully jailbreaks the Aristotle translation layer and engineers a
malicious Lean proof, the resulting execution remains permanently trapped within the WASM sandbox,
neutralizing the threat of systemic compromise~\cite{sandboxways2026}.

\section{Competitive and Market Landscape}

\subsection{Evaluating Existing AI Guardrails: Probabilistic vs.\ Formal Approaches}

As organizations race to operationalize generative AI, a vibrant ecosystem of guardrail solutions has
emerged. However, a critical comparative analysis reveals that the dominant market players—NVIDIA
NeMo Guardrails and Guardrails AI—rely on methodologies that are fundamentally misaligned with the
deterministic requirements of financial regulation.

\textbf{NVIDIA NeMo Guardrails} is a comprehensive, open-source toolkit that utilizes a specialized
modeling language called Colang to define dialog flows and topical
boundaries~\cite{nvidiablog2026trustworthy}. The system architecture is inherently
probabilistic~\cite{nvidiaghub}. When a user or agent submits a prompt, NeMo invokes an LLM to generate
a ``canonical form'' of the intent, then utilizes vector similarity algorithms to match this intent against
predefined rules~\cite{nvidiacolang2026}. This approach introduces significant vulnerabilities: vector
spaces are continuous, and adversarial perturbations can easily manipulate the semantic distance,
causing the system to misclassify a restricted action as benign~\cite{arxivagentssecurity2025}.
Furthermore, NeMo relies heavily on secondary ``LLM-as-a-judge'' calls for evaluation, which incurs
severe computational latency—often adding hundreds of milliseconds to the critical path—and decreases
overall system throughput~\cite{nvidiaperformance2026}.

\textbf{Guardrails AI} focuses on syntactic validation, utilizing the RAIL specification and
Python-based libraries (like Pydantic) to enforce structural consistency on LLM
outputs~\cite{ajayverma2026guardrails}. While syntactic validation is fast and deterministic, it is
logically shallow~\cite{aiagentstore}. Pydantic can guarantee that an agent's proposed trade volume is
an integer, but it cannot dynamically prove that the specific integer, when combined with historical
margin usage and real-time capital constraints, adheres to a multi-tiered regulatory policy.

The Lean-Agent Protocol transcends both paradigms. By shifting the verification burden to the Lean~4
kernel, the platform abandons probabilistic vector matching (NeMo) in favor of deep mathematical
theorem proving, while simultaneously vastly exceeding the logical depth of simple schema validation
(Guardrails AI)~\cite{aiagentstore}. The architecture provides absolute, binary certainty of compliance
with ultra-low, microsecond latency, uniquely positioning it as the only viable solution for
high-frequency, regulated financial execution~\cite{awslean2026}.

\begin{table*}[t]
\caption{Comparative Analysis of AI Guardrail Platforms}
\label{tab:comparison}
\setlength{\tabcolsep}{8pt}
\renewcommand{\arraystretch}{1.4}
\begin{tabularx}{\textwidth}{%
  >{\bfseries}p{3.0cm}   
  X                        
  X                        
  X                        
}
\toprule
  \normalfont\textbf{Feature}
  & \textbf{NVIDIA NeMo Guardrails}
  & \textbf{Guardrails AI}
  & \textbf{Lean-Agent Protocol (Aristotle)} \\
\midrule

Core Architecture
  & Colang dialog flows, vector similarity search, LLM
    prompts~\cite{nvidiacolang2026}
  & RAIL specification, Pydantic schema
    validation~\cite{ajayverma2026guardrails}
  & Aristotle auto-formalization engine +
    Lean~4 kernel~\cite{aristotle2025arxiv} \\

\addlinespace[4pt]

Verification Paradigm
  & \textit{Probabilistic} — semantic intent matching and
    LLM-as-a-judge evaluation~\cite{nvidiacolang2026}
  & \textit{Syntactic} — structural type-checking and
    regex constraints~\cite{ajayverma2026guardrails}
  & \textit{Deterministic} — Calculus of Inductive
    Constructions; binary proof/refutation~\cite{lean4lean2024} \\

\addlinespace[4pt]

Latency Profile
  & \textbf{High} — secondary LLM inference adds
    ${\sim}500$\,ms or more per call~\cite{nvidiaperformance2026}
  & \textbf{Low} — fast Python-level execution; no
    inference overhead
  & \textbf{Ultra-low} — kernel type-checking only;
    ${\sim}5\,\mu$s per verification~\cite{awslean2026} \\

\addlinespace[4pt]

Regulatory Fit \newline (SEC Rule 15c3-5)
  & \textcolor{red}{\textbf{Inadequate}} — continuous vector
    space vulnerable to adversarial drift and
    bypass~\cite{finra2025marketaccess}
  & \textcolor{red}{\textbf{Inadequate}} — cannot dynamically
    verify complex multi-variable capital
    logic~\cite{ajayverma2026guardrails}
  & \textcolor{teal}{\textbf{Optimal}} — immutable Lean~4 axioms
    provide mathematical certainty of policy
    adherence~\cite{finra2025marketaccess} \\

\bottomrule
\end{tabularx}
\end{table*}

\subsection{Applied Formal Methods in FinTech and Access Control}

The theoretical proposition of utilizing Lean~4 as an enterprise-grade execution gateway is powerfully
substantiated by existing large-scale deployments in cloud computing and decentralized finance.

The most prominent validation is found in Amazon Web Services' development of Cedar, an open-source
authorization policy language~\cite{awslean2026}. AWS utilizes Lean~4 in a process known as
``verification-guided development'' to formally model and verify Cedar's core components—the
evaluator, authorizer, and validator~\cite{awslean2026}. By proving overarching theorems such as
guaranteeing that a ``forbid'' policy will always trump a ``permit'' policy regardless of evaluation
order, AWS ensures that its access control mechanisms behave flawlessly across quadrillions of
production authorizations~\cite{awssecurity2023}. The efficiency of the Lean model—capable of
executing differential testing inputs in an average of 5 microseconds—proves unequivocally that formal
verification can scale to support the highest-throughput enterprise infrastructures in the
world~\cite{awslean2026}.

Furthermore, researchers have successfully leveraged Lean~4 to formalize the complex economic
behaviors of Automated Market Makers (AMMs) like Uniswap~\cite{ammLean2024}. Historically, the
economic security of decentralized exchanges relied on pen-and-paper proofs regarding constant-product
formulas ($x \cdot y = k$) and arbitrage opportunities~\cite{ammfee2026}. Recent studies have
mechanically formalized these properties, along with complex trading fee mechanisms, directly into the
Lean~4 theorem prover~\cite{ammLean2024}.

The critical importance of moving from theoretical models to machine-checked code is underscored by
landmark failures in algorithmic trading. The Knight Capital failure—where an unchecked algorithmic
error caused a \$440 million loss—and the 2007 ``Hammer'' case—where basic algorithms were
manipulated to illegally move settlement prices in NYMEX futures—demonstrate that financial markets
cannot tolerate software ambiguity~\cite{nasdaq2025algofailures}.

\section{Strategic Implementation Roadmap}

The deployment of the Lean-Agent Protocol within a financial institution requires a meticulous, phased
integration to manage the complexities of neuro-symbolic formalization and execution sandboxing.

\subsection{Phase 1: Minimum Viable Product — Stateless Policy Translation and Shadow Verification}

The objective of the MVP phase is to validate the auto-formalization capabilities of the Aristotle model
without exposing production systems to risk.

\begin{itemize}
  \item \textbf{Action:} Compliance teams input a static, bounded subset of institutional policies—such
  as basic capital thresholds derived from SEC Rule 15c3-5 and fundamental data privacy
  rules—into the Aristotle API.
  \item \textbf{Process:} Aristotle translates these natural language directives into Lean~4 axioms. To
  prevent formalization drift, engineering teams implement MenTaL-style concept-to-symbol
  constraints, ensuring rigorous, one-to-one mapping of financial terminology to Lean logic.
  \item \textbf{Execution:} The system operates asynchronously in ``shadow mode.'' Live agentic outputs
  are logged and subsequently passed to the Orchestrator for post-execution verification. The
  Lean kernel's judgments are compared against human compliance reviews to measure
  Aristotle's translation fidelity and baseline error rates.
\end{itemize}

\subsection{Phase 2: Beta — WASM Integration and Synchronous Interception}

Phase 2 transitions the protocol from an analytical tool into an active, synchronous security gateway.

\begin{itemize}
  \item \textbf{Action:} The verified Lean~4 Policy Environment is compiled and deployed within a
  hardened WebAssembly (WASM) runtime. WASI capabilities are strictly configured, granting
  the sandbox zero access to the host network or filesystem.
  \item \textbf{Process:} The Orchestrator node is moved into the critical path of the agentic loop. It
  intercepts real-time API requests, formulates Lean conjectures, and leverages SMT-backed
  \texttt{grind} tactics to achieve sub-millisecond kernel verification.
  \item \textbf{Execution:} The ``Right to Explanation'' back-translation pipeline is introduced.
  Utilizing the Herald dataset methodology, any action refuted by the Lean kernel triggers a
  reverse auto-formalization process. A localized RAG model translates the specific Lean error
  trace back into a plain-language adverse action notice, satisfying ECOA and FCRA mandates.
\end{itemize}

\subsection{Phase 3: Production-Ready — High-Frequency Deployment and Enterprise Scaling}

The final phase expands the Lean-Agent Protocol to encompass fully autonomous, high-frequency
enterprise operations.

\begin{itemize}
  \item \textbf{Action:} The Policy Environment is scaled to ingest complex, multi-tiered compliance
  rules spanning global directives (e.g., GDPR, EU AI Act, Basel III).
  \item \textbf{Process:} Advanced orchestration primitives, including bounded MPSC
  (Multi-Producer, Single-Consumer) mailboxes and work-stealing schedulers, are deployed to
  support concurrent multi-agent meshes, maintaining ultra-low latency under massive load.
  \item \textbf{Execution:} Comprehensive, cryptographically secure audit logging is finalized. Every
  agentic action executed by the firm is inextricably linked to an immutable Lean~4 mathematical
  proof stored in the execution telemetry. The institution achieves a zero-trust, mathematically
  verified AI infrastructure, enabling the limitless scaling of Agentic AI while remaining
  absolutely impervious to regulatory violations.
\end{itemize}

\section{Conclusion}

This paper has presented the Lean-Agent Protocol, a novel formal-verification-based guardrail
architecture designed to address the fundamental incompatibility between the probabilistic nature of
agentic AI and the deterministic demands of financial regulation. By leveraging the Aristotle
neural-symbolic model to auto-formalize regulatory mandates into Lean~4 axioms—and enforcing
execution through a Lean~4 kernel operating at microsecond latency—the protocol provides
cryptographic-level compliance certainty for the first time.

The protocol directly maps to and satisfies the requirements of SEC Rule 15c3-5 (Market Access Rule),
OCC Bulletin 2011-12 (Model Risk Management), FINRA Rule 3110 (Supervision), ECOA/FCRA
adverse action mandates, and EU AI Act explainability requirements. Comparatively, the protocol
transcends the probabilistic limitations of NVIDIA NeMo Guardrails and the logical shallowness of
Guardrails AI. Real-world deployments of Lean~4 at AWS (Cedar) and in DeFi (AMM formalization)
substantiate the technical and enterprise-scale viability of the approach.

The three-phase roadmap provides a structured path from shadow verification MVP to a production-ready,
zero-trust mathematical AI infrastructure—one that enables financial institutions to deploy autonomous
agentic AI at scale while remaining absolutely impervious to regulatory violations.

\section*{Acknowledgments}

V.~Rashi led the formal mapping of SEC Rule 15c3-5 and FINRA Rule 3110 regulatory mandates into the
Lean-Agent logical framework, providing the regulatory analysis that underpins
Sections~\ref{sec:sec15c3} and~\ref{sec:finra3110} of this paper.

\bibliographystyle{ACM-Reference-Format}
\bibliography{lean_agent_refs}

\end{document}